\documentstyle[12pt]{article}
\date{ }
\begin{document}
\title{Incoherent scattering of light by a Bose--Einstein 
condensate of interacting atoms}
\author{I.E.Mazets,\\A.F.Ioffe Physico-Technical Institute, \\
194021, St.Petersburg, Russia. \\
{\small E-mail: mazets@astro.ioffe.rssi.ru}}
\maketitle

\begin{abstract}
We demonstrate that incoherent photon 
scattering by a Bose--Einstein condensate of non-ideal 
atomic gas is enhanced due to bosonic stimulation of 
spontaneous emission, similarly to coherent scattering in 
forward direction. Necessary initial population of  
non-condensate states is provided by quantum depletion of 
a condensate caused by interatomic repulsion.   
\end{abstract}
 
\vskip 5pt


\vskip 5pt

Since the most reliable methods of diagnostics of a 
Bose--Einstein condensate (BEC) of a dilute atomic gas are 
based on laser spectroscopic techniques, the optical 
properties of BEC are of geat interest.  One of the most 
important features of the interaction of BEC with 
resonant light is the 
enhancement of  scattering in the forward direction 
\cite{JJ,You}. If an excited atom returns, after 
spontaneous emission of a photon, to the state occupied by a 
large number $N$ of atoms, the probability of such a process 
is increased in proportion to $N+1$ by the effects of quantum 
Bose--Einstein statistics. A degenerate atomic sample 
interacts with external light as an integral object, so  
this process is coherent. 
Because of momentum conservation, photons are 
scattered mostly into the forward direction. The 
corresponding solid angle $\Omega _{coh}$ 
is small but finite, because the 
atomic cloud size $R$ is finite, and is of order of 
$(kR)^{-2}$, where $k$ is the resonant photon wavenumber. 
 
On the other hand, it is commonly believed, that the 
photon scattering into modes lying outside of $\Omega _{coh}$, 
that leads to an escape of atoms from the condensate and 
heating of the atomic sample, is not modified with 
respect to the usual single-atom case \cite{You, LKP}. But 
such a conclusion is valid for an ideal gas only. 
In the present paper, we show how repulsive interaction of 
atoms composing BEC modifies incoherent photon scattering. 

Let us define, firstly, the initial $\left| i\right \rangle $
and final $\left| f\right \rangle $ states of the system. 
Initially, we have BEC of atoms in their internal ground state 
(the BEC chemical potential is $\mu $), no 
elementary excitation is present. As in Refs.\cite{JJ,You,LKP}, 
we consider a zero-temperature case. One atom is optically 
excited due to photon absorption and moves with respect to 
the BEC at the velocity ${\bf v}_{in}={\bf p}_{in}/m$, where 
$m$ is the mass of the atom, 
$\left| {\bf p}_{in}\right| =\hbar k$, and the $z$-axis of the 
reference frame is chosen to be parallel to ${\bf p}_{in}$. 
Then the atom undergoes spontaneous relaxation, but does not 
return to BEC, so the final state corresponds to the presence of 
one elementary excitation (quasiparticle) above the BEC. 

Now we can start with the formula for 
the probability per unit time to emit a photon with the momentum 
${\bf p}_{out}$ directed into the elementary solid angle 
$d\Omega =d\phi \,\sin \theta \, d\theta $ \cite{LL}: 
\begin{equation}
dw =\frac {\omega ^3}{2\pi \hbar c^3}\left| \left \langle f\left| 
{\bf e}^*\hat {\bf d}\right| i\right \rangle \right| ^2d\Omega .
\label{dw}
\end{equation}
Here the difference between energies of the initial and final 
states is equal to the energy $\hbar \omega $ of the emitted 
photon; {\bf e} is the polarization unit vector, $\hat {\bf d}$ 
is the dipole moment operator. Explicitely, we can write 
\begin{equation}
{\bf e}^*\hat {\bf d}=(
D_{eg}^*\hat a^\dag _{g\, {\bf q}}\hat a_{e\, {\bf p}_{in}}+
D_{eg}\hat a_{g\, {\bf q}}\hat a_{e\, {\bf p}_{in}}^\dag 
)P(\theta ). 
\label{ed}
\end{equation}  
$D_{eg}$ is the transition dipole moment matrix element, 
the operator 
$\hat a_{e\, {\bf p}_{in}}$ annihilates an optically excited 
atom with the momentum ${\bf p}_{in}$, 
the operator $\hat a^\dag _{g\, {\bf q}}$ creates an atom 
in the ground internal state with the momentum 
${\bf q}={\bf p}_{in}-{\bf p}_{out}$, $P(\theta )$ accounts 
for angular distribution of spontaneously emitted photons, 
$\left| P(\theta )\right| ^2= \frac 12 (1+\cos ^2\theta )$ 
for the circular polarization of the emitted photon.

The final state is not an eigenstate of the operator of number of  
particles with $q=0$ but an eigenstate of the 
quasiparticle operator number. The creation (annihilation) 
operators $\hat \alpha _{\bf q}^\dag $ 
($\hat \alpha _{\bf q}$) of a 
quasiparticle are introduced via the well-known Bogolyubov's 
canonical transformation which in the semiclassical 
approximation \cite{TD} reads as 
\begin{equation}
\hat a _{g\, {\bf q}}=\exp (-i\mu t/\hbar )\left[ 
u_q\hat \alpha _{\bf q}\exp (-i\epsilon (q)t/\hbar )+
v_q\hat \alpha _{\bf q}^\dag \exp (i\epsilon (q)t/\hbar )
\right] .               \label{BT}
\end{equation}
The transformation coefficients are 
\begin{equation}
u_q=\sqrt{\frac {\epsilon _{HF}(q)}{2\epsilon (q)}+\frac 12}   
,\qquad 
v_q=-\sqrt{\frac {\epsilon _{HF}(q)}{2\epsilon (q)}-\frac 12}. 
\label{uv} 
\end{equation}
Note that 
\begin{equation}
u^2_q+v^2_q=1.      \label{quadr} 
\end{equation}
In this context {\bf q} should be treated as the quasiparticle 
momentum. The quasiparticle energy is equal to 
$$
\epsilon (q)=\sqrt{q^4/(2m)^2+gn_cq^2/m},
$$
the Hartree-Fock energy is equal to 
$$
\epsilon _{HF}(q)=q^2/(2m)+gn_c  . 
$$
Here $n_c$ is the local density of the BEC (we use for it the 
Thomas--Fermi approximation \cite{TD}), $g=4\pi \hbar^2 a/m$ is the 
interatomic interaction constant, $a>0$ is the {\it s}-wave 
scattering length. 

Since, by definition,  
\begin{equation}
\left \langle f\left|\hat \alpha ^\dag _{\bf q} 
\hat a _{e\, {\bf p}_{in}}\right| i\right \rangle =1,\qquad 
\left \langle f\left|\hat \alpha _{\bf q} 
\hat a _{e\, {\bf p}_{in}}\right| i\right \rangle =0, 
\label{otk}
\end{equation} 
we get the following resulting formula 
\begin{equation}
dw=\frac {\omega ^3}{2\pi \hbar c^3}\left |P(\theta )\right| ^2
\left| D_{eg}\right| ^2 \left( 1+\bar v^2_q\right) d\Omega , 
\label{zn}
\end{equation}
where 
\begin{equation}
\bar v^2_q=\frac 1V\int d^3{\bf r} \, v^2_q ,
\label{vbar} 
\end{equation}
and the integral is taken over the volume $V=\frac 43\pi R^3$ 
occupied by the BEC. The spatial averaging given by 
Eq.(\ref{vbar}) appears when we pass from the local density 
treatment to consideration of a finite-size trap where 
$n_c({\bf r})$ is non-unform.  

Because $\epsilon (q)\ll \hbar \omega $, i.e. photon 
scattering can be regarded as an elastic process, we can apply 
the formula $q=2\hbar k\sin (\theta /2)$. 

Eq.(\ref{zn}) has a very transparent physical explanation. 
Indeed, $dw$ is the sum of the main part, which 
describes the spontaneous emission rate in the case of 
a single atom or in the case of vanishing interatomic 
interaction, and the positive correction term proportional to 
$\bar v^2_q$. But $v^2_q$ is the 
occupation number of an elementary cell 
$(2\pi \hbar )^3$ of the phase space surrounding the point 
({\bf r},~{\bf q}), and $\bar v^2_q$ is its spatially 
averaged value. Since this occupation number is non-zero, 
a bosonic stimulation of spontaneous relaxation takes place. 
Atoms with $q>0$ are present in the lowest 
energetic state of the system (the vacuum of 
quasiparticles) because of interatomic repulsion.  
This phenomenon is known as quantum depletion of a BEC. So we 
proove that 
it causes modification of incoherent light scattering. 

Let us discuss now a possibility to observe this effect in  
experiment. The total, integrated over angles, spontaneous 
emission rate differs very slightly from the standard value 
$2\gamma =\frac 43\hbar ^{-1}(\omega /c)^3\left| D_{eg}\right| ^2$, 
and in the first order with respect to the small parameter 
$\bar \beta =4\pi k^{-2}a\bar n_c$, where 
$\bar n_c =V^{-1}\int d^3{\bf r}\, n_c$ is the averaged 
condensate density, is equal to 
\begin{equation}
w=2\gamma \left( 1+\frac 38\bar \beta \right)  .    \label{totw}
\end{equation}
It is hard to detect such a small difference. So one should examine 
light scattering at small angles 
\begin{equation}
\theta \ll \sqrt{\bar \beta }.                       \label{an}
\end{equation}
In this range 
\begin{equation}
\frac {dw}{d\Omega }\approx \frac {3\gamma \sqrt{\bar \beta }\,
(V\sqrt{\bar n_c})^{-1}\int d^3{\bf r}\, \sqrt{n_c}}{8\pi \theta }
. \label{y1}
\end{equation}
This is much greater than the analogous value for forward light 
scattering by a single atom which is equal to $3\gamma /(8\pi )$. 
To provide a possibility to distinguish between this modified 
incoherent light scattering and the coherent 
one, the inequality 
\begin{equation}
\bar \beta \gg (kR)^{-2},      
\label{i1}
\end{equation}
has to be satisfied. Eq.(\ref{i1}) 
means that quantum depletion effects provide light scattering 
at relatively large angles in comparison with the coherent process 
caused by the BEC presence. Eq.(\ref{i1}) also can be written in the 
form $N\gg R/a$, which is the well-known condition application of  
the Thomas--Fermi approximation \cite{TD}. In any case, Eq.(\ref{y1}) 
holds for $\theta\, ^>_\sim \, (kR)^{-1}$, because the minimum 
value of $q$ in a finite-size trap is of order of $\hbar /R$.   

Such an additional incoherent scattering at angles of order of 
$\sqrt{\bar \beta }$ also can be explained as follows. Positions 
of the centers of mass of atoms in a BEC are not independent, but, 
due to interaction, a pair correlation is essential at distances 
up to $(4\pi an_c)^{-1}$ \cite{Kad}. 
Such a correlation, in other words, a small-scale atomic density 
inhomogeneity, results in 
additional scattering at angles given by Eq.(\ref{an}). A similar 
effect has been proposed recently \cite{SF} to detect a 
Bardeen--Cooper--Schriffer transition in a trapped fermionic 
$^6$Li gas. 

As a conclusion, we make some estimations for the parameters of 
the experiment of Ref.\cite{se}. A spherically symmetric trap 
contains $1.6\cdot 10^6$ sodium atoms (the scattering length 
$a = 2.75\cdot 10^{-7}$ cm, the resonant wavenumber 
$k=1.07\cdot10^5$ cm$^{-1}$), the cloud size 
$R=3.63\cdot 10^{-3}$ cm, so $\sqrt{\bar \beta }=0.049$ and 
$(kR)^{-1}=0.0026$, therefore additional incoherent forward 
scattering caused 
by quantum depletion of BEC can be easily distinguished from the 
coherent process studied in Refs.\cite{JJ,You}. 
Of course, such a cloud is optically dense 
for a radiation detuned from resonance less than to few 
hundreds MHz. But if the optical density is smaller than 
$\bar \beta ^{-1/2}$, then multiple photon scattering 
does not wash out the discussed effect. Namely, excess peak 
in angular distribution of incoherently scattered photons becomes 
less high and more wide but still noticeable. 
 
This work is supported by the Russian 
Foundation for Basic Researches (project No. 99--02--17076) 
and the state program ''Universities of Russia'' 
(project No. 990108).


\begin{thebibliography}{99}
\bibitem{JJ} J.Javanainen. Phys. Rev. Lett. {\bf 72}, 2375 (1994).

\bibitem{You} L.You, M.Lewenstein, R.J.Glauber, J.Cooper. 
Phys. Rev. A {\bf 53}, 329 (1996). 

\bibitem{LKP} U.Leonhardt, T.Kiss, P.Piwnicki. Eur. Phys. J. D 
{\bf 7}, 413 (1999). 

\bibitem{LL} V.B.Berestetskii, E.M.Lifshitz, L.P.Pitaevskii. 
''Quantum Electrodynamics''. Nauka, Moscow (1980).  

\bibitem{TD} S.Giorgini, L.P.Pitaevskii, S.Stringari. 
J. Low Temp. Phys. {\bf 109}, 309 (1997);  
L.P.Pitaevskii. Uspekhi Fiz. Nauk, {\bf 168}, 641 (1998); 
F.Dalfovo, S.Giorgini, L.P.Pitaevskii, S.Stringari. 
Rev. Mod. Phys. {\bf 71}, 463 (1999).  

\bibitem{Kad} B.B.Kadomtsev, M.B.Kadomtsev. Uspekhi Fiz. Nauk, 
{\bf 167}, 649 (1997). 

\bibitem{SF} W.Zhang, C.A.Sackett, R.G.Hulet. Phys. Rev. A 
{\bf 60}, 504 (1999).  

\bibitem{se} L.V.Hau, B.D.Busch, C.Liu, Z.Dutton, M.M.Burns, 
J.A.Golovchenko. Phys. Rev. A {\bf 58}, R54 (1998). 
\end{thebibliography}
\end{document}